\documentstyle[aps,preprint]{revtex}

\def\beq{\begin{equation}}
\def\eqn{\end{equation}\noindent}
\def\LEE{\Lambda}
\def\LE{$\LEE$~}
\def\LEc{$\LEE$,}
\def\LEp{$\LEE$.}
\begin{document}
\title{Curvature Fluctuations and the Lyapunov exponent at Melting}
\author{Vishal Mehra and Ramakrishna Ramaswamy}
\address{School of Physical Sciences, Jawaharlal Nehru University, New
Delhi 110067, India}
\date{\today}
\maketitle
\begin{abstract}

We calculate the maximal Lyapunov exponent in constant-energy molecular
dynamics simulations at the melting transition for finite clusters of 6
to 13 particles (model rare--gas and metallic systems) as well as for
bulk rare--gas solid. For clusters, the Lyapunov exponent generally
varies linearly with the total energy, but the {\it slope} changes
sharply at the melting transition. In the bulk system, melting
corresponds to a jump in the Lyapunov exponent, and this corresponds to
a singularity in the variance of the curvature of the potential energy
surface. In these systems there are two mechanisms of chaos -- local
instability and parametric instability. We calculate the contribution
of the parametric instability towards the chaoticity of these systems
using a recently proposed formalism. The contribution of parametric
instability is a continuous function of energy in small clusters but
not in the bulk where the melting corresponds to a decrease in this
quantity.  This implies that the melting in small clusters does not
lead to enhanced local instability.
\end{abstract}

\section{INTRODUCTION}

In recent years the dynamics of finite condensed matter systems,
especially atomic and molecular clusters, has been extensively studied
from a nonlinear dynamics perspective \cite{general}.  Various
quantities of interest like Lyapunov exponents, Lyapunov spectra,
distributions of finite-time Lyapunov exponents and the Kolmogorov
entropy have been computed to see evolution of chaoticity and
ergodicity. Very general considerations lead to the expectation that
the Lyapunov exponents and the Kolmogorov entropy should increase with
energy. However, there are indications that different systems can
display a variety of behaviour, and details of how such indices change
and the different kinds of possible (qualitative as well as
quantitative) behaviour---the various universality classes, so to
speak---have not yet been completely characterized.

In the present work we calculate the largest Lyapunov exponent,
\LEc for small rare-gas and metal clusters (modelled, respectively, by
the Lennard-Jones (LJ) and the many--body Gupta (Gu) \cite{g81}
potentials) as well as for bulk rare-gas solid, namely 256 LJ atoms in
a box with periodic boundary conditions. In the range studied, \LE is
generally linearly related to energy (except at very low energies) but
shows a sharp change in slope at an energy which can be related to the
melting transition.

We also compute an estimate for \LE through a semi-empirical
methodology provided by a recently proposed geometrical theory of
hamiltonian chaos \cite{clp96}. Under certain approximations this
allows for the estimation of the relative contribution of stable modes
of a hamiltonian system towards chaotic behaviour. The approximations
inherent in the theory \cite{clp96} are fulfilled in bulk systems
(where $N$ is large), but do not appear entirely satisfactory in small
clusters. This geometrical theory has as its input the curvature of the
configuration space manifold and its fluctuation.  These quantities and
their variation with temperature are themselves interesting because
they yield a statistical quantification of the potential energy
surface. Recent approaches to the computation of Lyapunov exponents
from (local) instantaneous mode analysis \cite{sastry,cr97} use this
information implicitly.

The simplest understanding of chaotic dynamics in such (hamiltonian)
systems is the standard KAM picture \cite{ll}. When the number of
freedoms becomes large (for 13 atoms, the phase space is 78
dimensional) the picture of a phase space foliated by tori, with
surrounding stochastic regions \cite{ll} is not particularly relevant.
However, the motion for specific initial conditions remains nearly
periodic, while for others there can be a positive Lyapunov exponent.
In particular the KAM theorem underestimates the chaotic thresholds by
several orders of magnitudes \cite{pl90} in high-dimensional systems as
the critical perturbation scales as $\sim \exp{-\alpha N_f}$ which
rapidly goes to zero with increasing $N_f$ (degrees of freedom),
contrary to the experience in numerical simulations beginning with
FPU's famous result \cite{fpuo}.  And attainment of chaoticity does not
exhaust the interest in dynamics---in particular the evolution of the
dynamics near a thermodynamic phase--transition is nontrivial.  At the
energies corresponding to the phase--transition phenomenon, the
accessible phase space increases greatly, and correspondingly
\LE shows a signature of the transition. An alternate means of analysis
is through decorrelation of the eigenvectors of the instantaneous
Jacobian along a trajectory \cite{sastry}.  The timescale for this
process is greatly reduced by the presence of the regions of negative
curvature (unstable modes) which also increase at the melting
phase--transition point.

Such ideas have been at the root of a variety of studies of the
Lyapunov exponent or related quantities.  Posch and Hoover
\cite{ph89} calculated Lyapunov spectra of solid and liquid LJ
systems in two and three dimensions, and attempted to fit a power law
to this data, obtaining different exponents in the solid and liquid
phases, although no definite significance could be ascribed to these.
Berry and coworkers \cite{hb93} have examined a variety of quantities
including finite-time or local approximations to Kolmogorov entropy and
Lyapunov spectrum in Lennard-Jones (LJ) and Morse clusters containing
between 3 and 13 particles \cite{hb93}. These studies have been able to
make contact between the features of the potential energy surface and
the variation of different dynamical indicators. Recently Sastry
\cite{sastry} has computed the maximal Lyapunov exponent as well as the
entire Lyapunov spectra for a 32 atom Lennard-Jones fluid in the
temperature range 50--800 K. A power law fit of \LE with temperature
does not yield a single exponent but a crossover (around $T=1$) between
the exponent = 1 at low temperature to 1/2 at high temperatures
\cite{sascomm}.  In a finite lattice system, Butera
and Caravati \cite{bc87} simulated the coupled rotor system in two
dimensions which has a Kosterlitz-Thouless (KT) transition at constant
energy and observed that the scaling of \LE with the temperature
changes at precisely the KT temperature. Recent simulations on the XY
model in 2 and 3 dimensions by Caiani {\it et. al.}
\cite{cccp97} have shown the differences between KT transition and true
symmetry--breaking transition in three dimensions.  The crossover in
scaling of \LE with temperature or energy per particle, also observed
in other lattice models where it had purely dynamical significance
\cite{fm94}, was suggested to coincide with the crossover
from slow to fast diffusion in the phase space \cite{pl90} and was
labelled as the strong stochasticity threshold (SST). The generality of
SST in nonlinear hamiltonian systems is not obvious, although it
appears to persist even for large degrees of freedom in the models
studied. It was also found that SST occurs in lattice models with
Lennard--Jones interactions in two and three dimensions \cite{bas96},
the signature being a rapid relaxation of the specific heat and
independence of \LE on the initial conditions.  Similar transitions
have also been observed in small rare-gas and metal clusters
\cite{mr97}. A somewhat different interpretation of these
results has also been proposed \cite{abcm96}.

In small rare-gas atomic clusters, \LE has been calculated at the
melting transition \cite{nrc1}, which is a finite-size analogue of the
bulk melting transition \cite{lw90}. Whereas
\LE changes discontinuously with energy for LJ$_{13}$
and LJ$_{55}$, for LJ$_7$, the slope changes discontinuously.  At the
energy of this change, indicators of melting like Lindemann index or
density of states show characteristic signature, so it was proposed
that \LE is a good indicator of melting transition \cite{nrc1,mnr97}.
Subsequent work 
\cite{mrk96} has revealed that the behaviour of \LE
as a function of the energy is more complicated and can be different
depending on the nature of the low energy configuration that the system
starts from.  More recently Dellago and Posch \cite{dp96} have
calculated \LEc the Lyapunov spectra and related quantities like the
fraction of unstable modes for the melting transition of modified LJ
systems in 2D (the so-called WCA and LJS potentials). They find that
\LE has a broad peak near the melting density and a steplike increase
at the melting temperature, and further suggest that there is a change
in the shape of the Lyapunov spectra at the transition density.
Critical phenomena which occur at higher temperatures in larger
fragmenting clusters have also been studied
\cite{blr95}, although the claimed form for the (finite-time) Lyapunov
exponent, namely \(\LEE\sim(T-T_c)^{-\mu}\), with universal behaviour
for $\mu$ is questionable \cite{cccp97,nbm96}.

Our work in the present paper is focussed on the study of clusters and
bulk at the melting transition, with particular emphasis on the
Lyapunov exponent and the curvature fluctuations. In the next section,
we examine the behaviour of \LEc and in Section 3, the nature of the
curvature fluctuations are analysed in order to make contact between
theory and simulations. The methodology and theoretical background for
the extraction of \LE from curvature data is elaborated in Section 3,
where our results for bulk as well as cluster systems are also
presented. This is followed by a summary and discussion in Section~4.

\section{MELTING AND LYAPUNOV EXPONENT IN CLUSTERS AND BULK}
\label{melt}
As is by now well--known \cite{ss89}, atomic clusters with as few as
six or seven particles undergo a finite size analogue of the bulk
melting phase transition, which is marked by a jump in the Lindemann
index and the onset of rapid isomerization. The liquidlike phase of the
small clusters is however unlike the bulk liquid in one important
respect: due to the phase space constraints, the atomic displacements
(``hopping processes'') are highly correlated, giving rise to $1/f$
spectra of single particle potential energy fluctuations \cite{nrc2}.
It has been observed that in metal clusters particles can occupy two
types of sites---of low and high energy respectively, and the onset of
the liquid phase corresponds to the onset of an isomerization occuring
through four particle interchange between high and low energy sites
\cite{b96}. Similar features are expected for rare--gas clusters also.
This peculiar dynamics has interesting consequences which we examine in
this section. 

We consider clusters of up to 13 atoms interacting via Lennard-Jones
(LJ) potential
\begin{equation} V = 4 \epsilon \sum_{i<j} \left(\left( \frac{ \sigma}{
r_{ij}}\right)^{12}-
\left(\frac{\sigma }{r_{ij}} \right) ^{6} \right) \end{equation}
\noindent
which is appropriate for rare--gas atoms; we work in reduced units when
$\sigma = \epsilon = 1$. In order to model metallic clusters
the manybody Gupta (Gu) potential \cite{g81}, 
\begin{eqnarray}
V = &(&1/2)U\sum_j\{A\sum_i\exp(-p(r_{ij}-r_0))\nonumber\\
&&-[\sum_i\exp(-2q(r_{ij}-r_0))]^{1/2}\}.
\end{eqnarray}\noindent
is commonly used. Here $r_{ij}$ is the distance between the atoms $i$
and $j$ and $r_0$ is the interatomic distance in the bulk (fcc)
crystal. The specific metallic cluster system we model corresponds
notionally to Ni, for which we use parameters given in Ref.\cite{ss89}:
$p=9/r_0, q=3/r_0, A=0.101035~\mbox{and~} U/E_{bulk}=0.85505$
($E_{bulk}$ is the bulk cohesive energy of the metal by which all
energies are scaled).  Quantities like \LE and Kolmogorov-Sinai entropy
have been calculated for small rare--gas clusters and the relation of
the potential energy surface to local dynamical behaviour has been
analysed \cite{hb93,ab93}.  In particular it is known that smooth
saddles cause a drop in local chaoticity indicators \cite{hb93}.

Time is measured in units of $(m\sigma^{2}/\epsilon)^{1/2}$ for LJ
clusters and $(mr_0^{2}/E_{bulk})^{1/2}$ for Gupta clusters.  Constant
energy simulations are done using the Verlet algorithm with stepsize
$\Delta t=0.01$ in these units, and the total energy is conserved to
$0.01\%$.  The total integration time varies between $10^{5}\Delta t$
to $5\times 10^{5}\Delta t$ depending on the potential and the system
size.  Simulations start at the global minimum of the structure and
then gradually move up in energy; at the highest temperatures studied
evaporation does not set in within the time period of the computation.
Temperature is defined in the usual manner, as proportional to the
average kinetic energy per particle, $T = 2\langle E_k \rangle/(k_B
(3N-6))$, $k_B$ being the Boltzmann constant.

At very low energies the dynamics is nonergodic, in particular for 13
particle clusters the nonergodicity can be persistent upto rather high
energies.  The breathing mode in LJ$_{13}$ was recently studied by
Salian {\it et. al.} \cite{sbr96}. This mode is stable ({\it i. e.~}
nondecaying) up to an energy $E=0.150$ per particle above the global
minimum; the corresponding mode for LJ$_{7}$ survives only up to energy
0.042 above the minimum. Stability here is tested by starting
trajectories with isotropic stretching of the global minimum structure.
In this nonergodic region the system initialised with small random
distortion of the icosahedral structure has a very small positive
Lyapunov exponent while for larger distortions at the same energy,
\LE can be much larger.  As energy increases, the
overall cluster distortions increase and this mode becomes markedly
unstable. For large initial distortions from icosahedral geometry of
LJ$_{13}$, a chaotic transition occurs at a lower energy and is
manifested as the divergence of the microcanonical specific heat.
\cite{mr97}. This energy depends on the 
equilibration time but tends to a nonzero value for large equilibration
times. Such transitions are not size--specific, and have been
seen for non--magic clusters as well.

Fig.~1 shows the variation of $\LEE(\epsilon)$ with the
reduced energy per particle, $\epsilon$, for LJ and Gu clusters of various
sizes. At higher energies $\LEE(\epsilon)$ is linear but
at a certain energy a sharp change in the slope is evident in all cases.
Precisely at this energy the conventional criterion of melting applies,
namely the Lindemann index crosses the value 0.1 (Lindemann indices for Gu
clusters have been studied in \cite{ss89}). In LJ$_{13}$ which has a large
solid--liquid coexistence region between $-2.6<\epsilon<-2.2$,
\LE changes slope at $\epsilon\approx-2.6$. In LJ$_{6}$
where the Lindemann index increases continuously, the discontinuity in
\LE appears just after the Lindemann index has reached
its critical value 0.1.  The slope in the liquid phase is always smaller,
indicating that the diffusive modes have different chaotic properties,
giving rise to different energy dependence of \LEp The
change in the slope of $\LEE(\epsilon)$ can also be taken
to be a characteristic signature of melting in small clusters.  The slope
of the $\LEE(\epsilon)$ curves are generally smaller for
the larger cluster; furthermore, the sharpness of the discontinuity (in
slope) is reduced as the cluster size increases.

For the Gupta clusters, however, these trends with $N$ are not
strongly--marked which is a consequence of the manybody character of the
Gu potential: even if a pair of particles is not interacting directly
(being outside the potential cutoff) the corresponding elements of the
Hessian matrix need not vanish because of the manybody term. We find that
the slope changes distinctly at an energy corresponding to top of the jump
in the Lindemann curve.

The third system we study is bulk, and Fig.~2(a) shows $\LEE(\epsilon)$
for the system of 256 LJ particles in a cubic box with periodic
boundary conditions at the reduced density $\rho=0.93$ and reduced
melting temperature 1.15.  Initial conditions for these simulations
were as follows: the atoms were initially at fcc lattice positions,
with initial velocities taken randomly from an appropriate gaussian
distribution.  For $\epsilon > -4.25 $ the crystal is unstable and soon
melts. It is possible for melting to occur for slightly lower energies
if integration is carried for longer times but the time required to
melt is not a monotonically decreasing function of energy. Here
$\LEE(\epsilon)$ shows a jump which obviously can be ascribed to the
increase in the disorder at melting.  The data shown is the Lyapunov
exponent of solid phase or the liquid phase trajectory only which are
obtained by discarding the premelting portion of a trajectory that
melts. Similar results for bulk melting have also been reported by
Dellago {\it et. al.} \cite{dp96} and Nayak {\it et. al.}
\cite{njbb96}.

While both the bulk and the cluster are disordered by melting (change
in entropy at melting, $\Delta S/N=1.0$ for LJ$_{55}$ and 1.7 for LJ
crystal \cite{lw90} and specific heat of even 6 particle cluster shows
a peak) \LE in the cluster liquid phase is significantly smaller than
the value obtained by extrapolating from the solid phase, in contrast
with
\LE obtained in bulk liquid.  This suggests that in the
cluster liquid phase there are stabilizing influences on the dynamics
which are absent in the bulk liquid. We conjecture that the correlated
hopping processes in clusters \cite{nrc2,b96} provide the necessary
mechanism.  This is consistent with the observation made above that the
$\LEE(\epsilon)$ curve gets smoother with the cluster size.

The observed Lyapunov phenomenon for the clusters is not just an effect of
the smeared--out bulk transition {\it i. e.~} a manifestation of finite
dynamical coexistence region in the clusters which vanishes in the bulk
limit. The properties of the coexistence region (in particular its
width) depend sensitively on the cluster geometry {\it e. g.~} the fact
that the cluster is magic or not. But the trends in the Lyapunov
exponent depend on the size in a simple manner and are independent of
the magic--nonmagic relation.

The spectra of $3N-7$ positive Lyapunov exponents for the clusters vary
smoothly with energy. For LJ clusters they are linear in the entire
range with a slope that increases with energy while in Gu clusters they
acquire increasing curvature. This is contrast to the results of
Dellago and Posch \cite{dp96} for bulk melting in two dimensions where
the curvature of the spectra changes sign smoothly at certain densities
(at constant temperature).  It remains a task to extract more useful
information from the shape of the Lyapunov spectra. However the
smoothness of the spectra at cluster melting implies that the relation
(if any) between thermodynamic and dynamic entropies is nontrivial.

\section{CURVATURE FLUCTUATIONS}

In this section we apply the geometric theory of Hamiltonian chaos
developed by Pettini and coworkers \cite{clp96} in order to interpret
and rationalize the results of our numerical simulations in terms of an
underlying (microscopic) mechanism. This theory, the salient features
of which are summarized below, is attractive because it attempts to
unite features of the potential energy surface with the dynamical
properties of the system as encoded in the Lyapunov exponents. One
additional motivation in applying this theory to finite cluster systems
is to determine the limits of applicability of the general framework,
which has mainly been applied to lattice models where the calculated
Lyapunov exponents are found to be in good agreement with empirical
exponents \cite{clp96,cccp97}.

\subsection{The Geometric theory of chaos in High-dimensional systems}

It is well known that the classical dynamics on manifolds of constant
negative curvature is chaotic \cite{gutz}. The dynamics on a manifold
with fluctuating positive curvature can also be chaotic \cite{p93}:
this fluctuation, via the mechanism of parametric instability, is
responsible for creating chaos in systems such as the
Fermi--Pasta--Ulam $\beta$ and $\phi^4$ chains \cite{pl90} and for a
driven one--dimensional oscillator studied by Chaudhuri {\it et. al.}
\cite{cgr93}. These studies have provided much of the motivation for
the development of the geometric theory of chaos \cite{clp96}.  Barnett
{\it et. al.} \cite{btnuf96} have applied similar ideas to calculation
of Lyapunov exponent of a dilute gas.

The geometric theory makes a diagonal approximation in the sense that
it uses information only about the trace of the instantaneous Hessian
matrix {\it i. e. ~} $\Delta V$. When the equations of motion are put
in a differential-geometric form, this term appears as the Ricci
curvature (curvature locally averaged over the directions) of the
enlarged configuration manifold in the Eisenhart metric \cite{clp96},
\beq
ds^{2}=-2V(q)dt^{2}+a_{ij}dq^{i}dq^{j}+dtdq^{N_f+1}. \eqn Here $V(q)$
is the potential energy and the kinetic energy is $T =
a_{ij}\dot{q}^{i}\dot{q}^{j}$.  The relevance of potential curvature
has been noted before \cite{toda}.  The Ricci curvature does not have
this simple form in other metrics but the essence of the assumption is
that all the directions in the phase space are equally curved after a
coarse-graining along a trajectory.  The dynamical trajectories are the
geodesics of this manifold.  In above models this appears to be the
dominant mechanism for chaos as there are no unstable modes
(corresponding to the negative eigenvalues of the instantaneous Hessian
matrix or regions of negative curvature) which are {\it local}
mechanism of chaos.

If it is assumed that the curvature fluctuations have gaussian spectra
upto a high-frequency cutoff {\it i. e.~} the dynamics generates a
gaussian random process for curvatures, then one can dispense with the
necessity of following a trajectory and an estimate of the largest
Lyapunov exponent, $\lambda$ can be obtained via Monte Carlo sampling.
This is the essence of the gaussian approximation \cite{clp96}, within
which excellent agreement is obtained between \LE and $\lambda$. The
presence of additional unstable modes renormalizes the calculated
exponent, although this is nontrivial to calculate.  Therefore the
unrenormalized exponent gives an estimate of the chaoticity caused by
stable modes only (the unstable modes also contribute to parametric
instability by their presence in $\Delta V$ but it is not their major
effect on chaoticity).  The theoretical basis of the diagonal
approximation assumption is that the local fluctuation of the Ricci
curvature detects deviation from isotropy (at a point) because
isotropic manifolds are necessarily of constant curvature by Schur's
theorem \cite{clp96}.

The crossover between the regimes of weak and strong chaos in
high-dimensional systems can be detected by examining the behaviour of
the mean curvature, $k$, as a function of the energy density. In the
integrable limit $k$ is independent of energy \cite{clp96}.
Corresponding to a Hamiltonian $H$ of $N$ particles with an interaction
$V$,
\begin{equation} H = \sum_{i}\frac{p_i^{2}}{2}+V(q)\end{equation}
there are $3N$ equations of motion,
\begin{equation}\frac{d^{2}q_i}{dt^2}=-\frac{\partial V}{\partial
q_i}.\end{equation} \noindent The associated Jacobi equation for the
second deviations is then
\begin{equation}\frac{d^{2}J_i}{dt^2}+\sum_j{\frac{\partial^{2}V}{\partial
q_i\partial q_j}J_{j}}=0,\end{equation} \noindent where $J_i$ are the
components of the vector of second deviations.  After some
approximations \cite{clp96} this can be converted to an equation for
$u=|J|$,
\begin{equation}
\label{oscillator}
\frac{d^{2}u}{dt^{2}}+\sum_{ij}{\frac{\partial^{2}V}{\partial q_i\partial
q_j}\frac{J_iJ_j}{|J|^2}}u\equiv \frac{d^{2}u}{dt^{2}}+Q(t)u=0
\end{equation}
which is, in effect, an equation for a linear oscillator with
time--dependent frequency $\sqrt{Q}$. The solutions of this equation
are unbounded for typical $Q(t)$ and the Lyapunov exponent is just
given by the rate of exponential growth of the envelope of $u$
\cite{kampen}. Pettini {\it et. al.} \cite{clp96} show that $Q(t)$ is
just the sectional curvature (which is the generalization of gaussian
curvature to many dimensions) relative to the plane containing $J$ and
${dq}/{dt}$ in the Eisenhart metric. The diagonal approximation
consists in replacing $Q$ by the simpler quantity
\begin{equation}
\label{simpler}
\frac{\Delta V}{N_f}=\frac{1}{N_{f}}\sum_i{\frac{\partial^{2}V}{\partial
q_i^{2}}}
\end{equation}
with $N_{f}$ the number of degrees of freedom, which is
\cite{clp96},
the Ricci curvature per degree of freedom {\it i. e.~} sectional curvature
has been averaged over relative orientations of $J$ and the velocity
vector.  Under the further assumption that the curvature is gaussian
distributed with a mean $k=\langle\negthinspace\Delta
V\negthinspace\rangle\negthinspace/N_{f}$ and variance
$\sigma^2=(\langle\negthinspace(\Delta
V)^{2}\negthinspace\rangle-\langle\negthinspace\Delta
V\negthinspace\rangle^{2})\negthinspace/N_{f}$ and are $\delta$-correlated,
Pettini {\it et. al.}
\cite{clp96}
derive an expression for an estimate of the Lyapunov exponent
\begin{eqnarray}
\label{parametric}
\lambda &=&\frac{1}{2}(l-\frac{4k}{3l})\\
l& = &\left(\sigma^{2}\tau+\sqrt{\frac{64k^{3}}{27} +
(\sigma^{2}\tau)^{2}}\right)^{1/3},
\end{eqnarray}
where $\tau$ is a characteristic time implied by the smoothness of the
underlying manifold {\it i. e.~} the time-interval below which dynamics of
curvatures cannot be regarded as a random process.

\subsection{Application of the Geometric theory}

The main result of the geometric theory is an estimate of the Lyapunov
exponent, $\lambda$ given in Eq.~(\ref{parametric}), for which it is
necessary to obtain the mean curvature, $k$ and the variance,
$\sigma^2$. These quantities can be calculated along a typical
trajectory using Eq.~(\ref{simpler}), and the assumption of
$\delta-$correlated curvature fluctuations can be directly verified.

The determination of the timescale $\tau$ (see Eq.~10) is
more tricky.  One estimate which has been used \cite{clp96,cccp97,cm96} is
\begin{equation}\label{tau}\tau=\frac{\pi\sqrt{k}}{2\sqrt{k(k+\sigma)}+
\pi\sigma}.\end{equation} \noindent
This expression for $\tau$ here is actually that for $\tau_{\star}$ in
Ref.\cite{clp96} (see the discussion following Eq.~45 there).  However
in the presence of negative curvatures it may be more accurate to
use a different timescale \cite{clp96}
\begin{equation}
\tau_2=\frac{k^{1/2}}{\sigma}
\end{equation}
We find (see the next subsection) that $\tau_2$ is more accurate than
$\tau$ insofar as it provides a better numerical match with the
autocorrelation decay timescale for the systems studied here.

One additional minor point is that the effective number of freedoms is
$N_{f}=3N-6$ rather than $3N$, since the six conserved quantities
(linear and angular momenta) give rise to zero frequency modes and thus
do not contribute to the chaoticity. This does not change qualitative
conclusions (indeed it must not) and improves results in most cases.

The application of the geometric theory to the systems considered in
Sec.\ref{melt} is of interest for two reasons. Firstly, this formalism
has so far been mostly applied to lattice models, where parametric
instability is the main source of chaos. It would be useful to
determine the extent to which the formalism works for off--lattice
models with significant local instability. Secondly, the partition of
the chaoticity of the system into local and nonlocal components may
help in clarifying the behaviour of Lyapunov exponent at melting. In
particular, it would be interesting to know whether the overall
instability of the system can be separated into these two components,
and if so, whether they behave differently at phase transitions.

The melting transition in finite systems appears to have a distinct
signature---the Lyapunov exponent shows a knee, where the slope changes
discontinuously \cite{mnr97}. For bulk, the fraction of the unstable
modes has a discontinuous jump at melting. Therefore a rapid increase
in local instability and consequently, a jump in the Lyapunov exponent
can be expected. Such a change may not be apparent in the contribution
of the parametric instability, and therefore in the situation of
cluster melting where Lyapunov exponent does not increase, it is an
open question whether the local instability increases or not.

\subsection{Results}

As should be clear from the preceeding discussion, application of the
geometric theory in order to make comparison with the results of our
numerical simulations involves a number of sensitive estimates and
several approximations.

Following the general procedure outlined in Sec.~IIIB above, we have
calculated the estimate $\lambda$ (cf. Eq.~\ref{parametric}) for bulk
(LJ) and various LJ and Gu clusters in an energy range which
encompasses melting, from long trajectories of duration up to $2\times
10^{6}\Delta t$. The data for $k,\sigma$ and $\lambda$ is shown in
Figs.~2-6. The detailed comparisons of theory and simulations are
presented separately for bulk and cluster systems below.

\subsubsection{Bulk LJ system}

Casetti and Macchi \cite{cm96} have calculated the curvature for bulk
LJ in a exponentially large energy range in order to detect the
crossover between weakly and strongly chaotic regimes. Earlier
calculations by LaViolette and Stillinger \cite{ls85} of the mean
curvature (which is proportional to the squared Einstein frequency)
show that $k$ increases linearly in the bulk LJ system with a jump of
about 20\% at melting.  This increase is a manifestation of the
positive high-frequency tail in the instantaneous normal spectrum in
the liquid phase.  However, the slope in the liquid phase is smaller
than the solid phase by about 6\%.

We find that the variance $\sigma^{2}$ has a {\it discontinuity} at an
energy $\epsilon_{m} =-4.17$.  At energies well away from
$\epsilon_{m}$, $\sigma$ increases linearly with energy but in a very
narrow range preceding $\epsilon_{m}$, roughly corresponding to the
solid--liquid coexistence region, $\sigma$ increases sharply. As the
system melts $\sigma$ falls by about 30\%. (The discontinuity in
$\sigma$ has been confirmed by repeating the calculations with longer
trajectories and finer energy mesh. Data in the coexistence region was
computed from long trajectories of total time $2\times 10^{5}\Delta t$.
Care was taken to ensure that computed averages are over either solid
or liquid phase exclusively.) This behaviour may be contrasted with the
cusp singularity found recently in XY model in three dimensions by
Caiani {\it et. al.} \cite{cccp97} which was interpreted as suggestive
of a topological transition in the potential energy surface.

The assumption of $\delta-$correlated curvature fluctuation can be
substantiated by examining the power spectra of curvature fluctuations.
Fig.~7 shows that these are satisfied in the bulk. However the observed
correlation time does not agree with $\tau$ given in Eq.~\ref{tau} (see
Table 1). We therefore use $\tau_2$ to calculate the relative
contribution of unstable modes, namely \beq
\delta\lambda_u \equiv
(\LEE-\lambda)/\LEE.
\eqn Indeed $\lambda(\tau_2)$ is much better fit to
\LE than $\lambda(\tau)$ (Fig 2).  $\delta\lambda_u$ can become slightly
negative in the solid phase (implying some correction due to
correlations), while in the liquid phase, it can be as large as 0.35.
Assignment of the difference $\LEE-\lambda$ to the effect of unstable
modes is justified by the following two observations. Firstly,
$\lambda$ does not increase at melting (it actually falls) whereas
\LE has significant jump which can accounted for by
an increase in the fraction of unstable modes..  In addition, the
agreement of $\lambda$ and \LE is better at lower temperatures, namely
when the occurrence of negative curvatures is infrequent.

\subsubsection{Clusters}

Owing to the finiteness of cluster systems, correlations do not decay
sufficiently rapidly \cite{nrc2}. As a consequence, curvature
fluctuations are far from being uncorrelated, and the framework of the
geometric theory breaks down for such systems and deviations from
Eq.~\ref{parametric} can be expected. The observed correlation times do
not agree with analytical estimates for $\tau$ and $\tau_2$ (Table 1).

The mean curvature for specific LJ clusters has been computed
previously \cite{as90,bm90}, and in contrast to bulk, decreases
uniformly with energy for all clusters (except LJ$_{13}$); likewise for
Gupta clusters.  No trend is apparent either for $k$ or its variation
with energy, although there are some size effects in the case of Gupta
clusters.  The behaviour of the variance, $\sigma^{2}$ is more complex.
Although this quantity usually increases smoothly with energy, in the
coexistence regime near melting there are large fluctuations which
persist for very long averaging times. The liquid phase in LJ$_{13}$
also shows nonmonotonic dependence of variance on energy. 

The net result of the persistence of correlations is that the estimates
$\lambda$ for cluster systems do not agree with \LEp  While
$\lambda(\tau)$ is a rather good fit for LJ clusters, it is of doubtful
validity because $\tau$ is far from the observed fluctuation timescale
$\tau_c$ (Table 1). Using $\tau_2$ in Eq.~\ref{parametric} gives
$\lambda/\LEE\approx 1.3-2.5$ for LJ clusters, although for the tightly
bound Gupta clusters, this discrepancy is smaller,
$\lambda/\LEE\approx0.8-1.2$.

\subsubsection{Discussion}

Our results indicate that unstable modes have suppressed chaoticity in
certain circumstances. In particular, in the solid LJ system, where
$\delta\lambda_u$ is very small, the fraction of unstable modes is
substantial ($\sim 0.2$) while slightly higher ($\sim 0.25-0.3$)
fraction of unstable modes in liquid gives greatly enhanced
$\delta\lambda_u~(\sim 0.35)$. In the cluster even after melting
$\delta\lambda_u$ does not increase very much: it has smooth dependence
on energy. A tentative conclusion that can be drawn from these cases is
that the unstable modes have effective chaoticity only when the
particles are free to execute large--scale motion.

If the coarse--grained curvature is everywhere positive, the ratio
$\sigma/k$ provides a crude measure of the ruggedness or roughness of
the underlying potential energy surface. As it is this feature which
causes nearby trajectories to diverge, it is interesting to study the
variation of this index with energy, as this will give some indication
of the nature of the region that is being dynamically probed on the
potential energy surface. 

At low energies $\sigma/k$ is small as expected, typical values being
$\sim 0.2-0.4$. At highest energies reached, it is between 0..8 and 1.2
for various clusters with somewhat higher values for LJ clusters and
smaller $N$.  In the bulk system, a  peak $\sigma/k\approx 0.8$ is reached
at the melting point (from the solid phase) and then it remains nearly
constant. The correspondence of the maximal roughness with
melting point is very suggestive. One can visualize destruction of the
crystal lattice being driven by large scale roughness of the potential.

\section{CONCLUSION}

In this paper we have examined the behaviour of the largest Lyapunov
exponent \LE as a function of energy in finite clusters of 6-13
rare--gas and metal atoms, and in bulk rare--gas solids.  These
systems undergo a phase transition from a regime wherein the dynamics
is purely oscillatory (involving individual particle vibrations) to a
regime where the dynamics is both oscillatory as well as diffusive.

Diffusive dynamics is linked to the presence of delocalized unstable
modes in the bulk \cite{bl95}.  In small clusters the onset of the
diffusion does not appear to enhance the chaoticity: the observed value
of the Lyapunov exponent is smaller compared to the value expected by a simple
extrapolation of the exponent from the low energy
regime,namely from the oscillatory dynamics or the ``solid'' phase.
This suppression of chaos, which we attribute to the correlated hopping
dynamics, is strongest for smallest clusters but is then progressively
reduced.  It is possible that for particular clusters, the enhancing
and suppressing effects of the unstable modes can balance and
$\LEE(\epsilon)$ curve is smooth across the melting (infact LJ$_{17}$
shows no signature of melting according to this measure
\cite{tbnm97}).  This conjecture can be tested by studies of the larger
clusters with calculations of the participation ratios of the unstable
modes which will clarify their role in the chaoticity of a dynamical system.

One may intuitively expect that unstable modes ({\it i. e.~} negative
curvatures) cause the dynamics to be chaotic. As noted by Dellago and
Posch in their study of melting in two dimensional systems\cite{dp96},
the fraction of unstable modes, which is a rough measure of negative
curvatures, has a similar dependence on the parameters as \LEp  These
unstable modes can however become important only when particles are
capable of large scale motion.  (In a related context, it has been seen
\cite{njbb96} that \LE falls when a liquid is cooled through its
glass--transition temperature, namely as the unstable modes get
localized \cite{bl95}.) 

Using the framework of a geometric theory of Hamiltonian chaos, we
compute an estimate for the Lyapunov exponent from the curvature of the
potential energy surface and its fluctuation. We have studied the
variation of these quantities with the temperature of the system and
found that the mean curvature is always a monotonic function of energy
but the variance has a simple energy dependence only for smaller
clusters. In the coexistence region of 13 particle clusters---these are
the cases in which potential energy surface has a deep global minimum
which is well separated from the next lowest structure---$\sigma$ is
nonmonotonic. The LJ bulk system shows singular behaviour for $\sigma$
at melting, which may indicate some sort of topological change in the
configuration space \cite{cccp97}.

The resulting estimate $\lambda$ is generally larger than \LE in the
solid phase. For bulk, the discrepancy is small, but for clusters, the
agreement is only qualitative. Curvature fluctuations for clusters are
correlated, and this effectively reduces the parametric instability in
the dynamics.  The spectral features of the curvature fluctuations such
as bandwidth are not well accounted for by the geometric theory even
for the bulk system.  At higher temperatures $\lambda$ is lower than
\LE which is attributed to
the unstable modes (negative curvatures) which are ignored in the
geometric theory. The contribution of the unstable modes towards
chaoticity (obtained by subtracting $\lambda$ from \LE and therefore
only approximate) is small in the solid phase but can be as large as
one--third in the liquid phase. Since $\sigma$ and $k$ do not show any
singularity at melting for clusters, the parametric contribution coming
from the change in topography of the potential energy surface is
changing smoothly.  Therefore, the fractional chaoticity coming from
the unstable modes, $\delta\lambda_u$, also seems to vary continuously
with energy.  

In summary, our application of the geometric theory to the dynamics of
the melting transition for cluster and bulk systems has provided a
satisfactory qualitative understanding of the underlying mechanisms in
terms of the change in roughness of the potential energy surface,
curvature fluctuations and and parametric instability.  While agreement
between theory and simulation is reasonable for the bulk system, for
the case of finite clusters, the situation is less satisfactory.  The
main source of the discrepancy seems to lie in the fact that in cluster
systems, correlations are temporally long lived.  This aspect requires
to be incorporated within the present framework of the geometric theory
(see {\it e. g.} Refs.~\cite{cgr93,kampen}) in order to achieve
quantitative accuracy. 

\section*{Acknowledgments:} We thank Srikanth Sastry for discussions,
generous advice, and for a critical reading of the manuscript. 
This work was supported by grant SPS/MO-5/92 from the Department of
Science and Technology.

\begin{figure}
\label{fig1}
\caption{(a) \LE as the function of reduced energy 
per particle for LJ
clusters with $N$=6,7,9,11,13. Note that the list includes both magic
and nonmagic clusters. (b) \LE for Gu clusters with 
$N$=6,7,13. The reduced energy scale is set by $E_{bulk}$ (Eq.~2)}
\end{figure}
\begin{figure}
\label{fig2}
\caption{(a) Lyapunov exponents for the bulk LJ system of 256 particles in a
cubical box with periodic boundary conditions. Circles refers to
\LE and squares to the estimate $\lambda$ generated
using Eq.~8 ~ with $\tau$ defined in Eq.~10. Pluses ($+$) denote values
of $\lambda$ calculated using $ \tau_2=k^{1/2} / \sigma $.  (b) Mean
curvature ($k$), and fluctuation $\sigma$ for bulk LJ system as a
function of energy.  $k$ and $\sigma$ are measured in units of
frequency squared.~}
\end{figure}
\begin{figure}
\label{fig3}
\caption{(a) Mean curvature $k$ and (b) fluctuation $\sigma$ for 
LJ$_{N}$ clusters with $N$=6, 11, 13 as a function of energy. Units are
as in Fig.~2.} 
\end{figure}
\begin{figure}
\label{fig4}
\caption{The estimate $\lambda$ for LJ$_{N}$ clusters with $N$=6, 11, 13
as the function of energy. Shown with circles are $\lambda$ calculated with
$\tau$ defined in Eq.~10; $+$ are values of $\lambda$ calculated using
$\tau_2=k^{1/2}/\sigma$. Also, for comparision, are the correspnding
values of \LE from Fig 1 (squares)}
\end{figure}
\begin{figure}
\label{fig5}
\caption{(a) Mean curvature ($k$), and (b) fluctuation ($\sigma$)
Gu$_N$ clusters with $N$=6, 7, 13 as
a function of energy. $k$ and $\sigma$ are scaled by $E_{bulk}/mr_0^{2}$.}
\end{figure}
\begin{figure}
\label{fig6}
\caption{The estimate $\lambda$ for Gu$_N$ clusters with $N$=6, 7, 13 as
a function of energy (same conventions as Fig.~4).}
\end{figure}
\begin{figure}
\label{fig7}
\caption{Power spectra of curvature distribution calculated along a
trajectory for (a) bulk LJ (solid), (b) bulk LJ (liquid), (c) LJ$_{13}$
at energy= -2.4 corresponding to the temperature 34K (coexistence
region) and, (d) Gu$_{13}$ at $\epsilon=-0.67$ (liquid phase). The
vertical axis is in arbitrary units.}
\end{figure}
\newpage
TABLE 1. Typical timescales (in reduced units) associated with power
spectra for various systems studied here.
\vspace{2cm}

\begin{tabular}{|c|c|c|c|c|c|}
\hline
System&~~~~ $\tau^{a}$ &~~~~ $\tau_2^{b}$ &~~~~ $\tau_B^{c}$
&~~~~$\tau_c^{d}$ &~~~~\LE\\ \hline Gu$_6$
solid&0.30&1.5&0.8&2.0&0.2~\\ \hline
~~~~~Liquid&0.18&0.4&0.7&1.1&0.7~\\ \hline Gu$_{13}$
solid&0.28&0.9&0.8&1.4&0.3~\\ \hline ~~~~~Liquid &0.18&0.36 &
0.7&0.9&0.8~\\ \hline LJ$_6$ solid &0.55&1.6&2.5&10.&0.05\\ \hline
~~~~~Liquid &0.4&0.8&1.0&1.0&.25\\ \hline LJ$_{13}$
solid&0.5&1.4&1.7&2.5&0.1\\ \hline ~~~~~Liquid&0.3&0.6&1.0&1.0&0.3\\
\hline
Bulk solid&0.2&0.5&0.6&0.6&0.3\\ \hline
~~~~~Liquid&0.2&0.6&0.4&0.4&0.8\\ \hline
\end{tabular}
\vskip1cm
\small
\noindent$^{a}$~~~calculated with Eq.~10.\\ $^{b}$~~~calculated with Eq.~11\\
$^{c}$~~~Obtained from approximate upper cutoff of the power
spectrum.\\$^{d}$~~~Inverse bandwidth of the power spectrum (approximate)\\

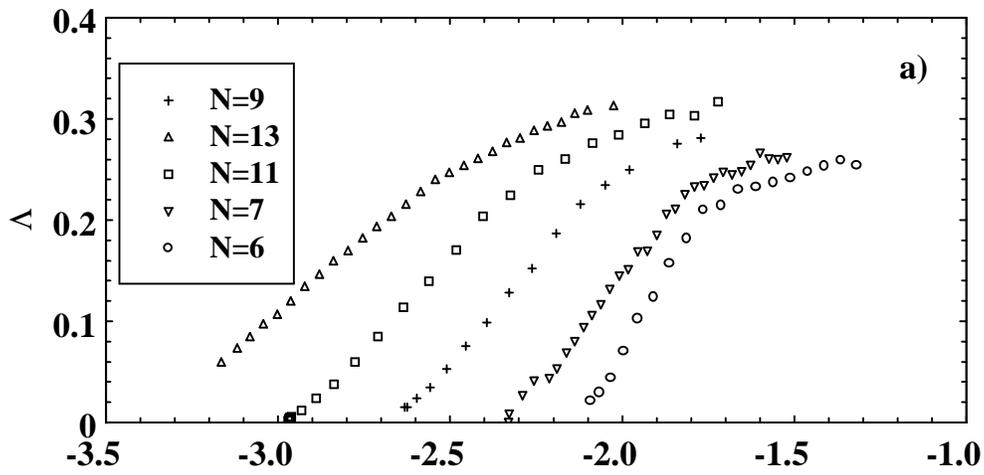
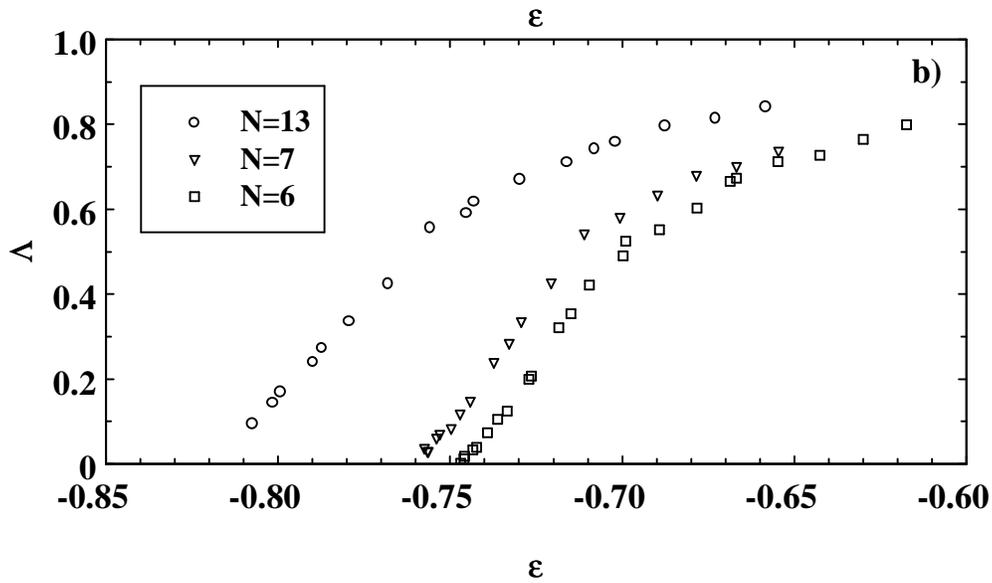
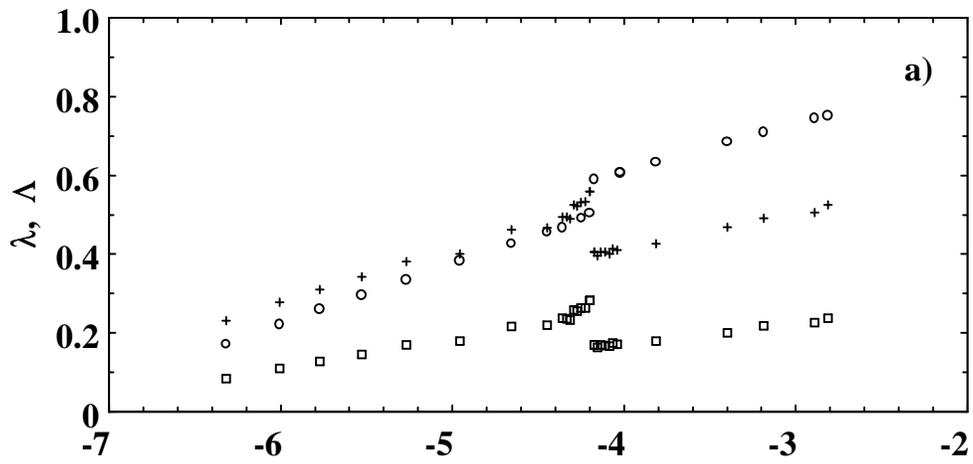
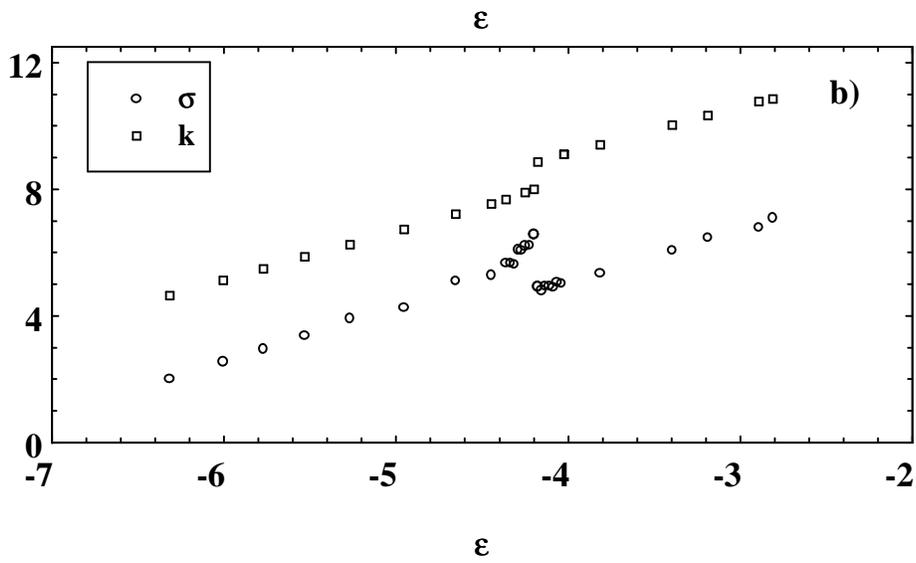
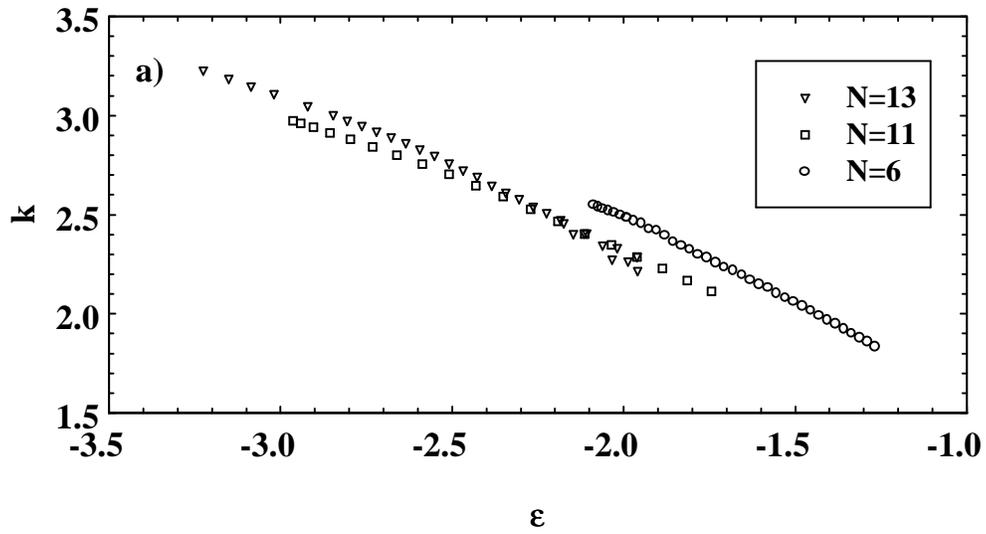
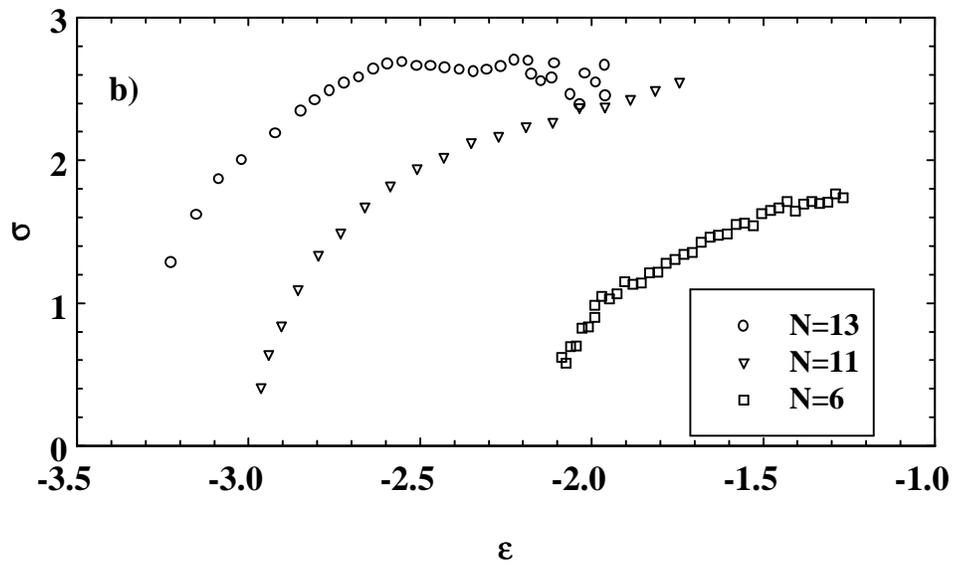
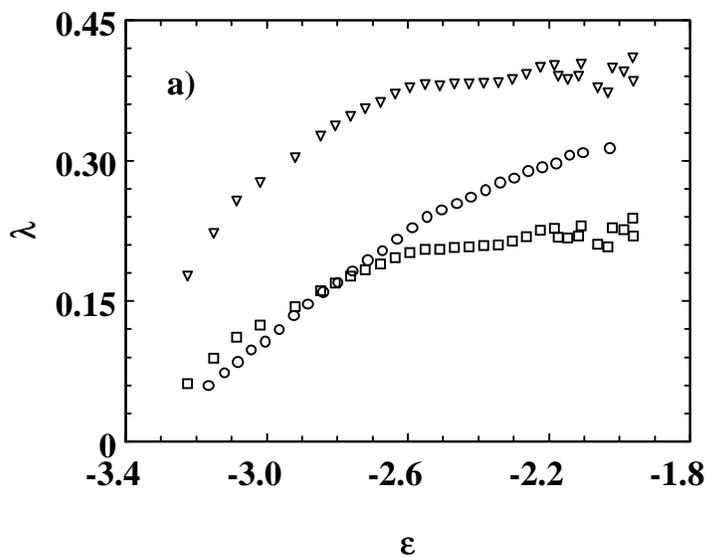
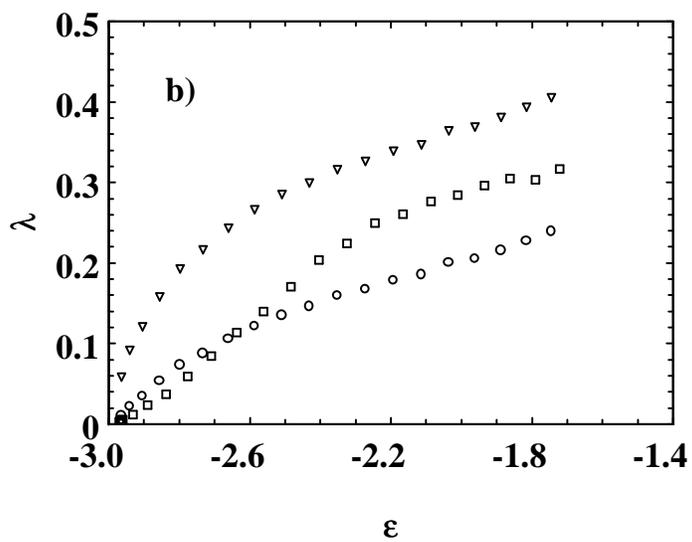
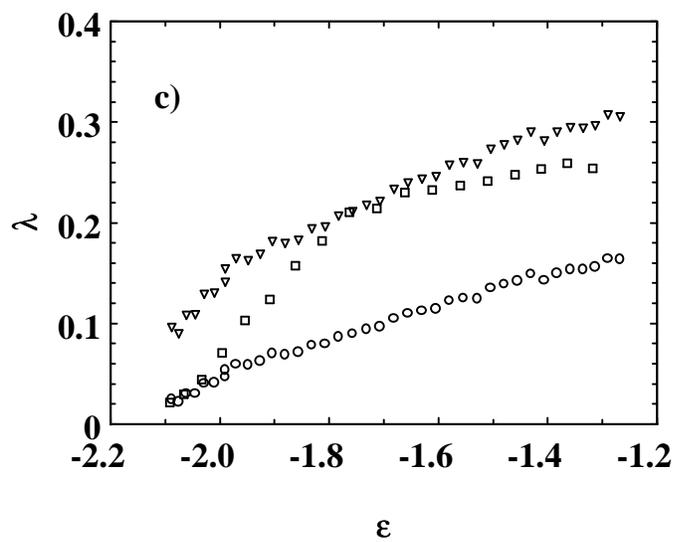
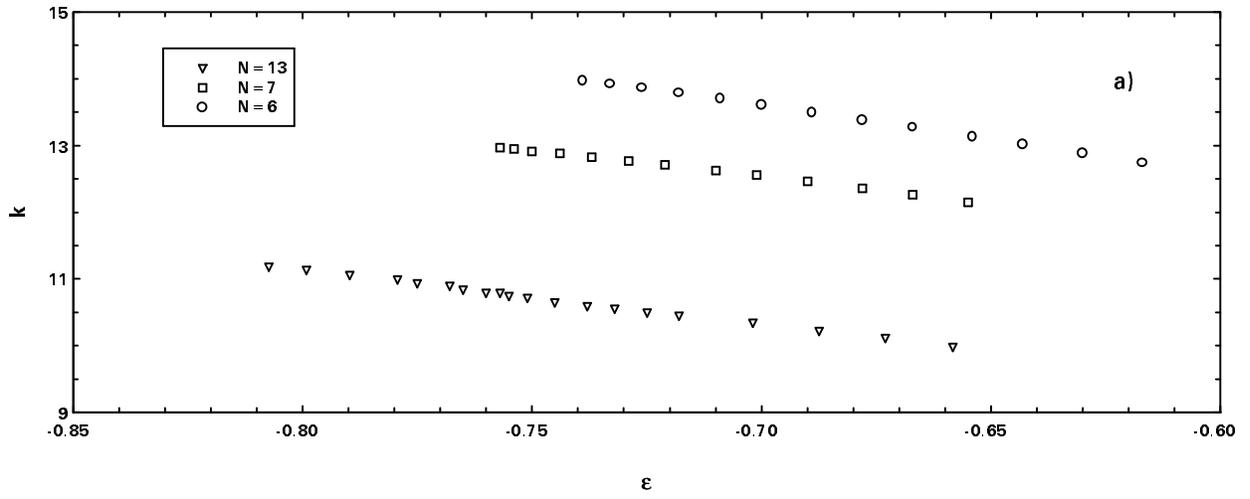
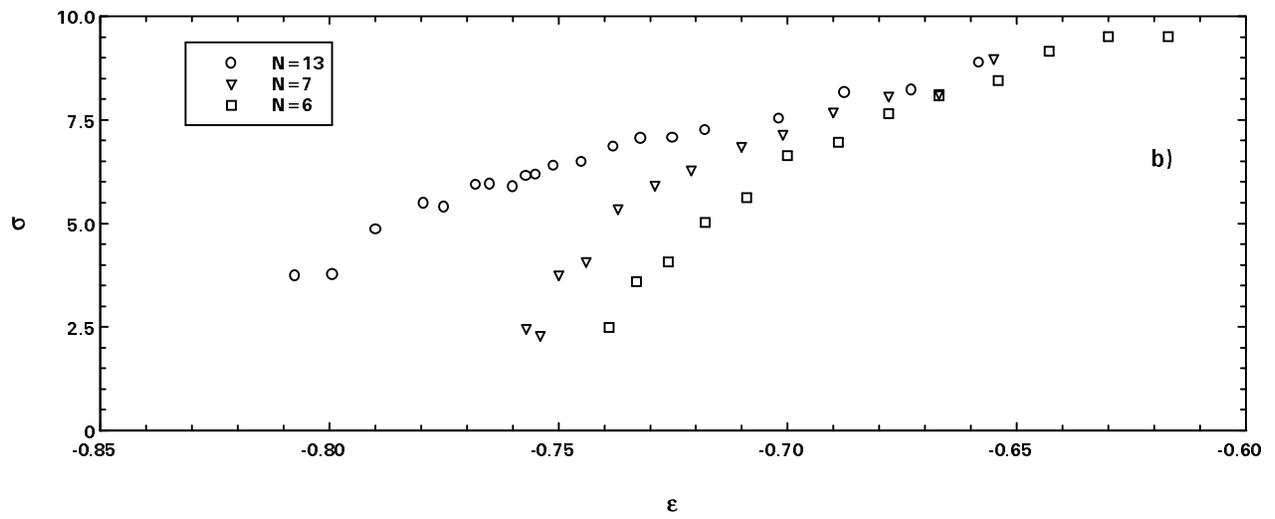
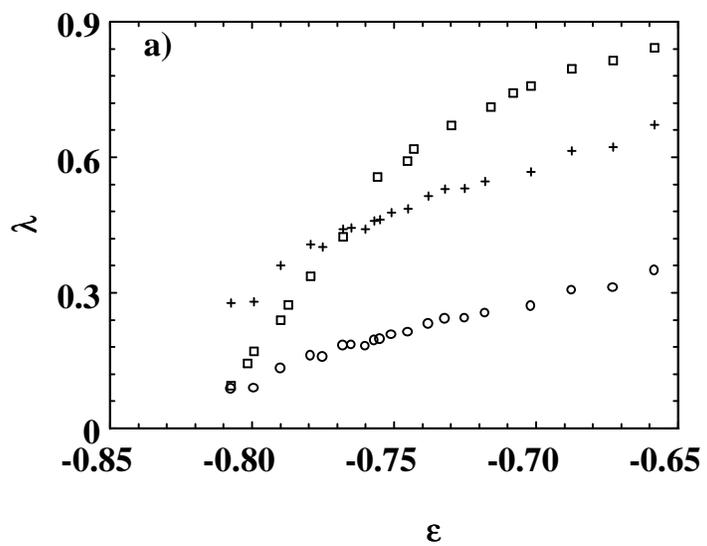
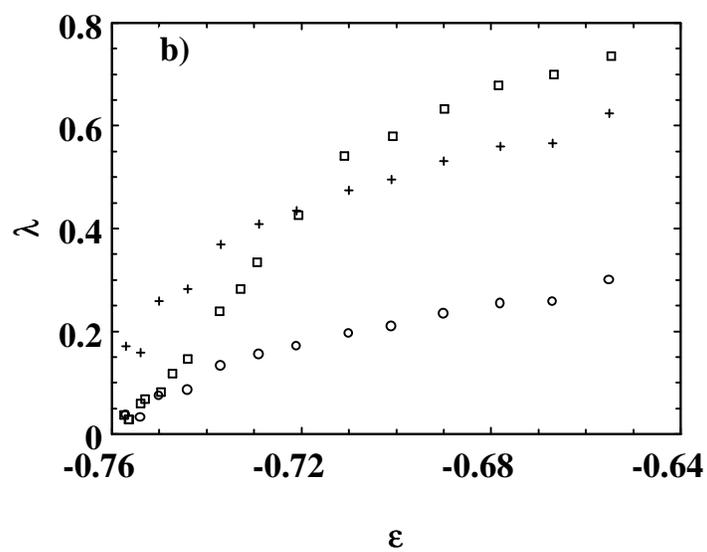
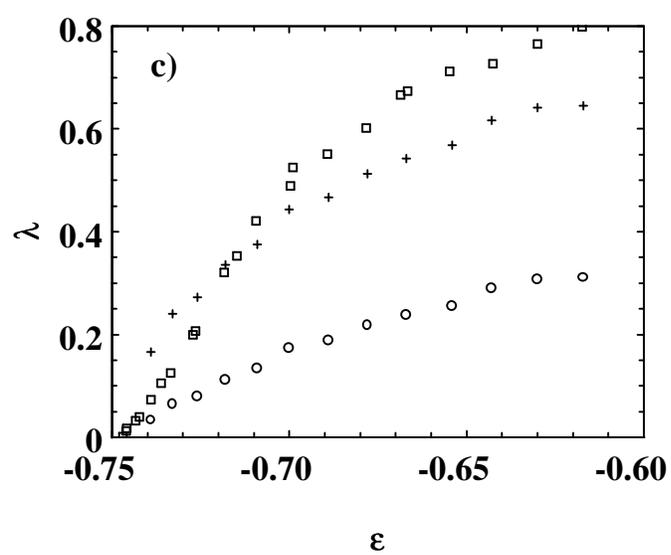
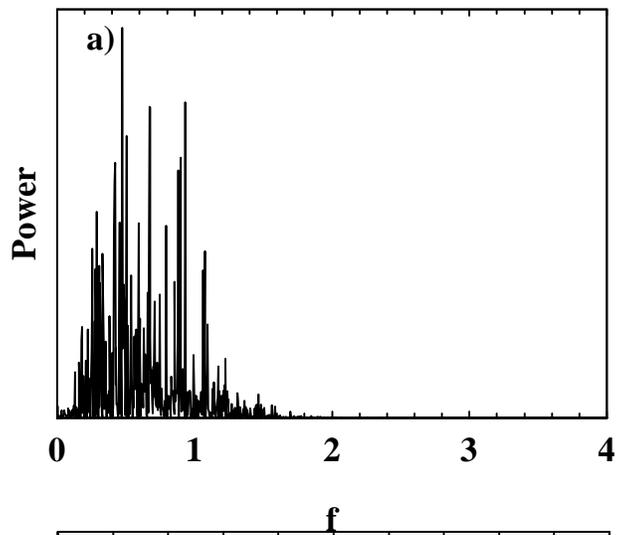
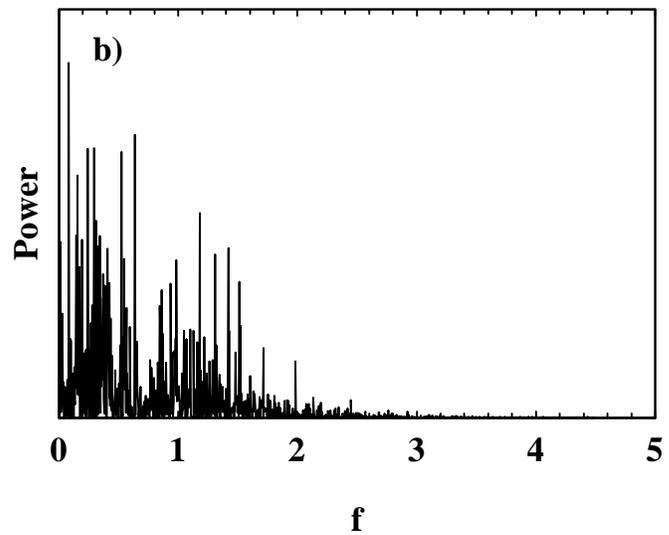
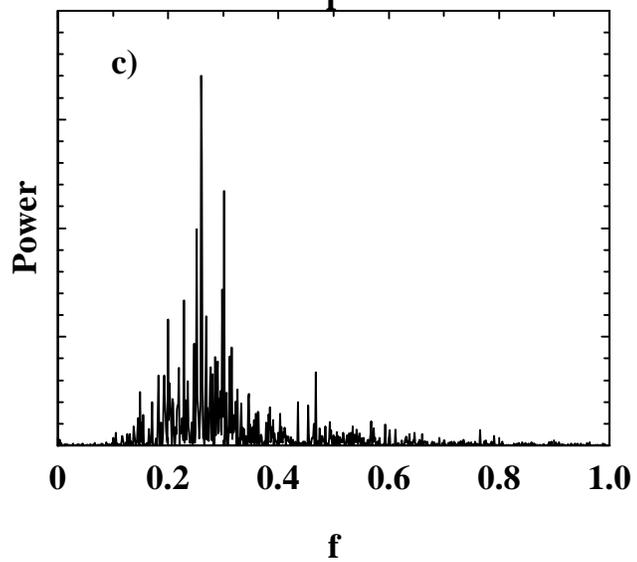
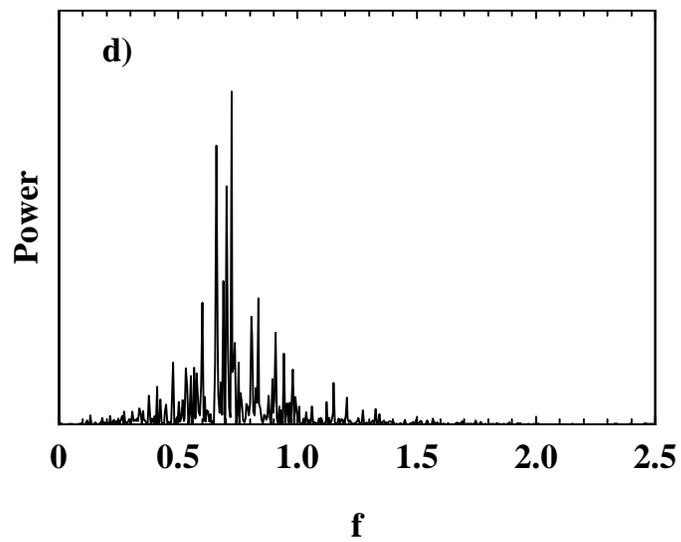
\end{document}